# Tuning selective reflection of light by surface anchoring in cholesteric cells with oblique helicoidal structures

OLENA S. IADLOVSKA[1,2], GRAHAM R. MAXWELL[1], GRETA BABAKHANOVA[1,3], GEORG H. MEHL[4], CHRISTOPHER WELCH[4], SERGIJ V. SHIYANOVSKII[1,3], AND OLEG D. LAVRENTOVICH[1,2,3,*]

**Affiliations:**

[1]*Liquid Crystal Institute, Kent State University, Kent, Ohio 44242, USA*

[2]*Department of Physics, Kent State University, Kent, Ohio 44242, USA*

[3]*Chemical Physics Interdisciplinary Program, Kent State University, Kent, Ohio 44242, USA*

[4]*Department of Chemistry, University of Hull, Hull HU6 7RX, UK*

*olavrent@kent.edu

## ABSTRACT

Selective reflection of light by oblique helicoidal cholesteric (Ch$_{OH}$) can be tuned in a very broad spectral range by an applied electric field. In this work, we demonstrate that the peak wavelength of the selective reflection can be controlled by surface alignment of the director in sandwich cells. The peak wavelength is blue-shifted when the surface alignment is perpendicular to the bounding plates and red-shifted when it is planar. The effect is explained by the electric field redistribution within the cell caused by spatially varying heliconical Ch$_{OH}$ structure. The observed phenomenon can be used in sensing applications.

## INTRODUCTION

Cholesteric liquid crystals (Ch) with periodically twisted arrangements of molecules are known as the prime example of the material with the so-called structural color [1, 2]. Since the period of the structure is often in the range of few hundreds of nanometers, Ch is capable of selective reflection of light in the visible part of spectrum. Recently it has been demonstrated that the peak wavelength and the band width of light reflection can be controlled by an electric [3-5] or a magnetic field [6]. The tunability is achieved when the material exhibits a small bend elastic constant [3, 7, 8]. Upon application of the field, such a material forms an oblique helicoidal structure, abbreviated as $Ch_{OH}$, in which the local director twists around the direction of the field and makes an acute angle $\theta$ with this direction. A non-zero projection of the director onto the field direction ensures a finite aligning field torque which controls the period of the structure. In a regular Ch, the molecules are perpendicular to the helicoidal axis and such a tunability is impossible.

Here we demonstrate that the peak wavelength and the bandwidth of selective reflection of light at the oblique helicoidal $Ch_{OH}$ can be controlled by surface alignment: the wavelength of Bragg reflection is blue-shifted when the alignment is perpendicular (homeotropic) and red-shifted when it is planar.

## RESULTS AND DISCUSSION

**Materials and measurement techniques.** We prepared a $Ch_{OH}$ mixture with a small bend constant $K_3$ at temperatures close to the room temperature. The components of the mixture are dimeric compounds 1″,7″ - bis(4-cyanobiphenyl-4′-yl) heptane (CB7CB) and 1″,11″- bis(4-cyanobiphenyl-4′-yl) undecane (CB11CB), rod-like mesogen pentylcyanobiphenyl (5CB) (Merck), and left-handed chiral dopant S811 (Merck), in weight proportion 5CB:CB7CB:CB11CB:S811 = 49.8:30:16:4.2. At $T_{Ch-I} = 60.7°C$, the Ch phase melts into an isotropic fluid, and at $T_{Ch-Ch_{tb}} = 24.5°C$ it transforms into the chiral analog of the twist-bend nematic [9-11]. The experiments were performed at $T = 27.5°C$. The temperature was controlled by a hot stage HCS402 with temperature controller mK2000 (both Instec, Inc.) with the accuracy $0.01°C$.

Planar cells were assembled from glass plates with transparent indium tin oxide (ITO) electrodes and a layer of rubbed polyimide PI2555. Homeotropic cells were assembled from ITO-coated glass plates spin-coated with a polyimide SE5661 (Nissan Chemical Industries) and then additionally coated with 1% water solution of Dimethyloctadecyl[3-(trimethoxysilyl)propyl] ammonium chloride (Sigma-Aldrich) [12]. The cell thicknesses were set by spherical spacers mixed with UV-curable glue NOA 68 (Norland Products, Inc.). The cell thickness was measured by the light interference technique. We report data for a planar cell with $d = 23.7$ μm and for a homeotropic cell with $d = 23.5$ μm.

An AC voltage of frequency 3 kHz (sinusoidal mode, DS345 waveform generator (Stanford Research) and 7602-M voltage amplifier (KROHN-HITE Co.)) is applied to the ITO electrodes. Since the mixture is of a positive dielectric anisotropy, the electric field causes formation of the heliconical director structure with the axis $\hat{\mathbf{t}}$ perpendicular to the glass plates and parallel to the field, $\hat{\mathbf{t}} \parallel \mathbf{E}$. The heliconical structure is stable in the range $(0.6 - 1.1)$ V/μm, changing its period with the field. When the field changes, the peak wavelength $\lambda_{max}$ of Bragg reflection moves from UV to near IR, as described in Refs. [3, 4]. The Bragg reflection is characterized by using USB2000 spectrometer and tungsten halogen light source LS-1 (both Ocean Optics). LS-1 generates a white unpolarized beam that is focused by the lens into a paraxial ray impinging normally onto the cells. $\lambda_{max}$ was determined by finding the full width of the reflection peak at its half-amplitude $\Delta\lambda$ and then defining $\lambda_{max}$ as the coordinate of the middle of the full width.

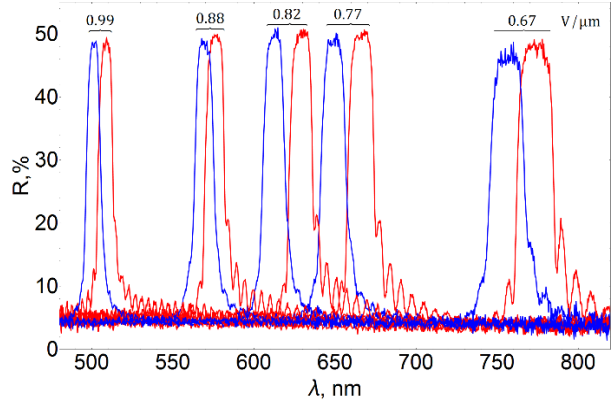

**Fig. 1.** (Color online.) Selective Bragg reflection in cells with homeotropic (blue) and planar (red) surface alignment under the applied electric field $E_{app} = U/d$. Values of $E_{app}$ are indicated above the reflection peaks in V/μm.

**Optical spectra results and discussion.** Figure 1 shows that the planar and homeotropic Ch$_{OH}$ cells of practically identical thickness $d$, being subject to the same electric field $E_{app} = U/d$, where $U$ is the voltage, produce different wavelength $\lambda_{max}$ of Bragg reflection. The homeotropic cell shows a blue-shifted reflection, while the planar cell shows a red-shifted reflection. The separation of the two peaks varies from 32 – 35 nm in the 750-800 nm region to 6-7 nm in the 500-550 nm region, Fig.1.

The field dependences of $\lambda_{max}$ for the homeotropic and planar cells are plotted in Fig. 2 (a). The difference is clear if the data are presented as $\lambda_{max} E_{app}$ vs. $E_{app}$, Fig. 2 (b). Although the planar cell shows wider peaks as compared to the homeotropic cell, this difference is small, Fig. 2 (c). The difference in spectral response is not related to the small difference in cell thicknesses, as the data in Figs. 1-2 are presented as functions of the field rather than the voltage.

To understand the mechanism behind the anchoring sensitivity of reflection, let us first recall the Ch$_{OH}$ features. The Ch$_{OH}$ appears under the action of electric field $\mathbf{E} = E\hat{\mathbf{z}}$ in order to minimize the free energy density [7, 13]:

$$f = \tfrac{1}{2}\left\{ K_1\left(\nabla\cdot\hat{\mathbf{n}}\right)^2 + K_2\left[\hat{\mathbf{n}}\cdot(\nabla\times\hat{\mathbf{n}}) - q_0\right]^2 + K_3\left[\hat{\mathbf{n}}\times(\nabla\times\hat{\mathbf{n}})\right]^2 - \mathbf{D}\cdot\mathbf{E}\right\}, \qquad (1)$$

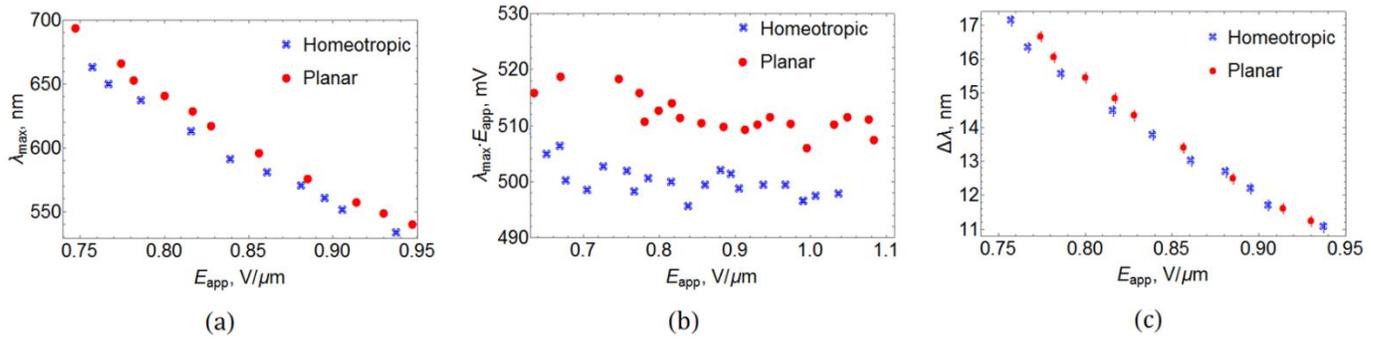

**Fig.2.** (Color online.) The wavelength $\lambda_{max}$ (a), product $\lambda_{max} E_{app}$ (b), and bandwidth $\Delta\lambda$ (c) of reflection peaks in homeotropic (blue) and planar (red) cells as functions of applied electric field $E_{app}$.

where $K_1$ and $K_2$ are the elastic constants of splay and twist, respectively; $q_0 = 2\pi/P_0$ is the field-free value of the wave-vector; the pitch $P_0$ is determined by the chemical composition; $\mathbf{D} = \varepsilon_0 \boldsymbol{\varepsilon} \mathbf{E}$ is the electric displacement, $\varepsilon_0$ is the dielectric constant and $\boldsymbol{\varepsilon}$ is the tensor of dielectric permittivity with the two principal components $\varepsilon_\parallel$ and $\varepsilon_\perp$, parallel and perpendicular to the director, respectively. For an infinitely thick Ch$_{OH}$ cell considered in Refs. [3, 7, 8], the equilibrium director writes in Cartesian coordinates $(x, y, z)$ as

$$n_x = \cos\varphi \sin\theta_\infty; \quad n_y = -\sin\varphi \sin\theta_\infty; \quad n_z = \cos\theta_\infty, \tag{2}$$

where $\varphi = qz + const$ is the angle of the director's azimuthal orientation, $q = 2\pi/P$; the pitch $P$ and the polar conical angle $\theta_\infty$ between the local director and $\hat{\mathbf{t}}$ are both field-dependent:

$$P = \frac{2\pi}{E}\sqrt{\frac{K_3}{\varepsilon_0 \Delta\varepsilon}}, \tag{3}$$

$$\sin^2\theta_\infty = \frac{\kappa}{1-\kappa}\left(\frac{E_{NC}}{E} - 1\right). \tag{4}$$

Here, $\Delta\varepsilon = \varepsilon_\parallel - \varepsilon_\perp$, $\kappa = K_3/K_2$ and $E_{NC} = 2\pi K_2/P_0\sqrt{\varepsilon_0 \Delta\varepsilon K_3}$ is the field at which the oblique helicoid unwinds into a uniform uniaxial nematic.

The peak wavelength of reflection at an infinitely thick Ch$_{OH}$ cell is $\lambda_{max}^\infty = \bar{n}_{eff} P$, where $\bar{n}_{eff} = (n_o + n_{e,eff})/2$ is the average refractive index, and $n_{e,eff} = n_o n_e / \sqrt{n_e^2 \cos^2\theta_\infty + n_o^2 \sin^2\theta_\infty}$ is the effective extraordinary index. With Eq. (3), the field dependency writes

$$\lambda_{max}^\infty = \frac{2\pi \bar{n}_{eff}}{E}\sqrt{\frac{K_3}{\varepsilon_0 \Delta\varepsilon}}. \tag{5}$$

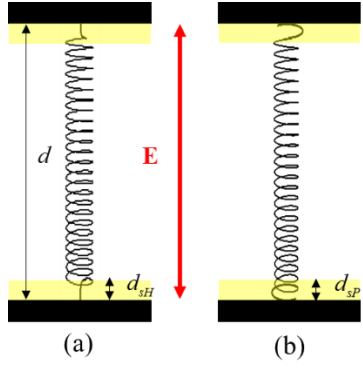

**Fig.3.** Schemes of Ch$_{OH}$ cells with director deformation in surface regions for homeotropic (a) and planar (b) alignment.

The bandwidth of reflection is $\Delta\lambda = \Delta n_{eff} \cdot P$, where $\Delta n_{eff} = n_{e,eff} - n_o \approx (n_e^2 - n_o^2) n_o \sin^2 \theta_\infty / 2 n_e^2$. Using Eqs. (3), (4) and the definition of $E_{NC}$ above, one concludes that the bandwidth is also tunable by the field:

$$\Delta\lambda_\infty = \frac{\kappa^2 E_{NC} P_0 n_o}{2E(1-\kappa)} \left(1 - \frac{n_o^2}{n_e^2}\right)\left(\frac{E_{NC}}{E} - 1\right). \tag{6}$$

Figure 3 illustrates how the heliconical Ch$_{OH}$ is influenced by the surface anchoring in cells. In the homeotropic cell, the conical angle $\theta(z)$ near the bounding plates depends on the vertical coordinate z and remains smaller than $\theta_{bulk}$, Fig.3 (a). In the planar cell, the subsurface angle $\theta(z)$ is larger than $\theta_{bulk}$, Fig. 3 (b). The structure in the center of cells is similar to the uniform oblique helicoid. The spatial variation $\theta(z)$ implies a variation of dielectric permittivity and thus, spatial variation of the local electric field. To describe the field distribution, we consider the cells as being comprised of three regions. In the homeotropic cell, there are two subsurface regions of a small thickness $d_{sH}$ in which $\theta$ changes from 0 at the substrates to some value $\theta_{bulkH}$, and a central region of thickness $d - 2d_{sH}$, where $\theta = \theta_{bulkH} = const$. In all three regions, the electric displacement along the z-axis, $D_z = \varepsilon_0 \varepsilon_{zz}(z) E_z(z) = const$, is independent of the z-coordinate; here $\varepsilon_{zz}(z) = \varepsilon_\perp \sin^2 \theta(z) + \varepsilon_\parallel \cos^2 \theta(z)$ and $E_z(z)$ is the spatially varying electric field. The model allows us to present the applied voltage through $D_z$:

$$U = 2d_{sH} \frac{D_z}{\varepsilon_0 \varepsilon_{sH}} + (d - 2d_{sH}) \frac{D_z}{\varepsilon_0 \varepsilon_{bulkH}}, \tag{7}$$

where $\varepsilon_{sH} = (\varepsilon_{bulkH} + \varepsilon_\parallel)/2$ is the effective permittivity of the two subsurface regions, $\varepsilon_{bulkH}$ is the permittivity of the central region with $\varepsilon_{bulkH} = \varepsilon_\perp \sin^2 \theta_{bulkH} + \varepsilon_\parallel \cos^2 \theta_{bulkH}$.

The value of $d_{sH}$ is much smaller than $d$ and determined by the elastic response of the helicoid to the surface alignment. Since the heliconical structure in the surface regions is not well defined, the measured selective reflection of light is determined mostly by the reflection from the central part with $\theta_{bulkH} = \text{const}$. The peak wavelength of this reflection can be described similarly to Eq.(5) in which $E = E_{bulkH} = D_z / \varepsilon_0 \varepsilon_{bulk}$. The peak $\lambda_{\max H}$ of reflection at a homeotropic cell of finite thickness can be written as a product of $\lambda_{\max}^\infty$ and two correcting factors,

$$\lambda_{\max H} = \lambda_{\max}^\infty \eta_H \xi_H, \quad \eta_H = \frac{\bar{n}_{bulkH}}{\bar{n}_{eff}}, \quad \xi_H = 1 - 2\frac{d_{sH}}{d}\left(1 - \frac{\varepsilon_{bulkH}}{\varepsilon_{sH}}\right); \tag{8}$$

here $\bar{n}_{bulkH} = n_o/2 + n_o n_e /2\sqrt{n_e^2 + (n_o^2 - n_e^2)\sin^2\theta_{bulkH}}$. Both correcting factors are smaller than 1. In Eq. (8), $\varepsilon_{bulkH} < \varepsilon_{sH}$, where $\varepsilon_{sH}$ is approaching $\varepsilon_\parallel$ because of the surface alignment; therefore, $\xi_H < 1$. Furthermore, $\theta_{bulkH} < \theta_\infty$, thus $\bar{n}_{bulkH} < \bar{n}_{eff}$ and $\eta_H < 1$. We conclude that both $\xi_H$ and $\eta_H$ favor a blue-shifted reflection of light in a homeotropic cell as compared to an infinite slab of Ch$_{OH}$, i.e., $\lambda_{\max H} < \lambda_{\max}^\infty$.

The peak wavelength $\lambda_{\max P}$ of the planar cell is derived similarly, by introducing two surface regions $d_{sP}$, in which the cone angle varies from $\pi/2$ to $\theta_{bulkP} = \text{const}$, so that

$$\lambda_{\max P} = \lambda_{\max}^\infty \eta_P \xi_P, \quad \eta_P = \frac{\bar{n}_{bulkP}}{\bar{n}_{eff}}, \quad \xi_P = 1 - 2\frac{d_{sP}}{d}\left(1 - \frac{\varepsilon_{bulkP}}{\varepsilon_{sP}}\right) \tag{9}$$

and $\varepsilon_{bulkP} = \varepsilon_\perp \sin^2\theta_{bulkP} + \varepsilon_\parallel \cos^2\theta_{bulkP}$, $\bar{n}_{bulkP} = n_o/2 + n_o n_e /2\sqrt{n_e^2 + (n_o^2 - n_e^2)\sin^2\theta_{bulkP}}$. Since $\theta_{bulkP} > \theta_\infty$ and $\varepsilon_{bulkP} > \varepsilon_{sP}$, one finds that $\eta_P > 1$ and $\xi_P > 1$, i.e. the reflection peak in a planar cell shifts towards the red end, $\lambda_{\max P} > \lambda_{\max}^\infty$, as observed.

In theory, the bandwidth $\Delta\lambda = \Delta n_{eff} \cdot P$ depends on $P$ and on $\Delta n_{eff}$; its experimental value is also influenced by nonuniformity of the textures. The difference in $\Delta\lambda$ between the homeotropic and planar cells is noticeable in the red part of spectrum, corresponding to the fields below 0.85 V/μm, Fig. 2 (c). At higher fields, there is practically no difference. The effect can be caused by a relatively large separation of $\lambda_{\max H}$ and $\lambda_{\max P}$ in the red part of spectrum that yields a larger difference in the pitch. Nonuniformity of the textures is another factor making $\Delta\lambda$ for the two cells similar to each other.

The model explains the main experimental result, namely, blue and red shifts of Bragg reflection in homeotropic and planar cell. It has a clear physical meaning. In the homeotropic cell,

7redistribution of the electric field is such that the bulk layer with a uniform Ch$_{OH}$ is subject to an electric field that is higher than the average field $U/d$. In the planar cell, the effective field is weaker than $U/d$. Since the pitch is inversely proportional to the field, Eq.(3), the shift is blue in the homeotropic cell and red in the planar cell.

It is of interest to estimate $d_s$ for the typical $\varepsilon_\perp, \varepsilon_\parallel, \varepsilon_{bulk}$. Assuming $d_{sH} \approx d_{sP}$, and $\varepsilon_{bulkH} \approx \varepsilon_{bulkP}$, one finds

$$\frac{d_s}{d} = c(\theta) \cdot \frac{\lambda_{\max P} - \lambda_{\max H}}{\lambda_{\max H}}, \tag{10}$$

where $c(\theta) = (\varepsilon_{bulk} + \varepsilon_\parallel)(\varepsilon_{bulk} + \varepsilon_\perp)/4(\varepsilon_\parallel - \varepsilon_\perp)\varepsilon_{bulk}$ is a weak function of $\theta$. With the permittivities measured in our laboratory, $\varepsilon_\parallel = 14.5$ and $\varepsilon_\perp = 6.4$, one finds $c(\theta) = 1.27 \pm 0.02$ when $0 < \theta < \pi/6$. Thus, $d_s$ is about 0.4 μm in the region 500 – 600 nm and 1 μm in the region 700-800 nm, i.e., much smaller than the cell thickness.

## SUMMARY

We demonstrated a pronounced effect of confinement and alignment on Bragg reflection from the Ch$_{OH}$ structure. The peak wavelength is red-shifted in planar Ch$_{OH}$ cells and blue-shifted in homeotropic cells. The shifts are explained by redistribution of the electric field within the cell, caused by different director orientation in surface regions and nonlocal character of the dielectric coupling. The sensitivity of Bragg reflection to the surface alignment can be used as a tool to detect chemicals that are capable of changing the anchoring direction in liquid crystal cells, see, for example [14].

**Funding.** National Science Foundation grant DMR-1410378 and REU program CHE-1659571.

**Acknowledgement**. We thank Dr. T.J. White, Air Force Research Laboratory, Wright-Patterson Air Force Base, Ohio, for the supply of CB7CB.